\documentclass[letterpaper, 10pt, conference]{ieeeconf}

\usepackage{subfiles}

\usepackage{algorithm2e}
\usepackage{graphicx}
\usepackage{amssymb}
\usepackage{amsmath}
\usepackage{cases}
\usepackage{siunitx}
\usepackage{tcolorbox}
\usepackage{lineno}
\usepackage{lipsum}
\usepackage{float}
\usepackage{caption}
\usepackage{geometry}

\newcommand{\figref}[1]{
    Fig.~\ref{#1}
}
\newcommand{\tableref}[1]{
    Table~\ref{#1}
}

\makeatletter
\newcommand{\removelatexerror}{\let\@latex@error\@gobble}
\makeatother
\newcommand\PromptContent[1]{``\texttt{#1}''}

\newcommand\Plant{\mathcal{P}}
\newcommand\User{\mathcal{H}}
\newcommand\Interpreter{\mathcal{A}}
\newcommand\Controller[1][\theta]{\mathcal{C}_#1}

\newcommand\midmid{~\middle |~}
\newcommand\Diag[1]{\mathrm{diag}\{#1\}}
\newcommand\InRange[2]{\in\{#1,\dots, #2\}}
\newcommand\Unit[1]{~\mathrm{#1}}

\newcommand\xDim{n}
\newcommand\uDim{m}
\newcommand\zDim{l}
\newcommand\thetaDim{q}
\newcommand\eDim{M}
\newcommand\XDim{N_p\xDim}
\newcommand\UDim{N_p\uDim}

\newcommand\RealField[1]{\mathbb{R}^{#1}}
\newcommand\xField{\RealField{\xDim}}
\newcommand\uField{\RealField{\uDim}}
\newcommand\zField{\RealField{\zDim}}
\newcommand\thetaField{\RealField{\thetaDim}}
\newcommand\XField{\RealField{\XDim}}
\newcommand\UField{\RealField{\UDim}}

\newcommand\InRealField[1]{\in\RealField{#1}}

\newcommand\Idx[1]{^{(#1)}}
\newcommand\IdxJ{\Idx{j}}
\newcommand\vase{\texttt{vase}}
\newcommand\toy{\texttt{toy}}

\newcommand\fint{f_\mathrm{int}}
\newcommand\fup{f_\mathrm{up}}

\newcommand\XSet{\mathcal{X}_\theta}
\newcommand\USet{\mathcal{U}_\theta}

\title{ChatMPC: Natural Language based MPC Personalization}

\author{Yuya Miyaoka, Masaki Inoue, and Tomotaka Nii}

\date{}
\begin{document}

\maketitle

\begin{abstract}
    We address the personalization of control systems, which is an attempt to adjust inherent safety and other essential control performance based on each user's personal preferences. 
    A typical approach to personalization requires a substantial amount of user feedback and data collection, which may result in a burden on users. Moreover, it might be challenging to collect data in real-time.
    To overcome this drawback, we propose a natural language-based personalization, which places a comparatively lighter burden on users and enables the personalization system to collect data in real-time.
    In particular, we consider model predictive control (MPC) and introduce an approach that updates the control specification using chat within the MPC framework, namely ChatMPC.
    In the numerical experiment, we simulated an autonomous robot equipped with ChatMPC.
    The result shows that the specification in robot control is updated by providing natural language-based chats, which generate different behaviors. 
\end{abstract}

\section{Introduction}

Defining the control specification is the primary concern in the designing of control systems. Specifications such as the control objective, and safety constraints are determined by a few people with domain knowledge and skills, aiming at the control system's robustness.
As usual, once the specification is decided, it will be fixed. All users would use the control system with the same specification, although there are various environments and user preferences.

To enhance the performance in each environment, it is necessary to incorporate the capability of adaption into the control system. For example, \cite{Bujarbaruah18}, \cite{Lorenzen19}, and \cite{Tanaskovic13} propose a method to estimate the state transition equation or transfer function of the plant from the input and output data of the plant, and reflect on the plant model in model predictive control (MPC).
Also, \cite{Liang22} and \cite{Parwana2022} propose a method to update the control barrier function (CBF) based on collected data. In \cite{Liang22}, CBF is expressed in the affine form and the optimal CBF is generated by learning its coefficients depending on the user's safety report. In \cite{Parwana2022}, an agent observes the movements of other agents to calculate the degree of danger to itself, then the CBF is updated by the degree of danger.

Adapting specifications is not limited to environmental factors. Recently, there is also specification updating according to user preferences, which is called ``personalization''. In the context of personalization, the specifications of the control system are updated so that the system suits each user's preference. 
For example, \cite{Zhu2021}, \cite{Maccarini22}, and \cite{Forogione20} propose a method to personalize the MPC. In the method proposed in \cite{Zhu2021}, \cite{Maccarini22}, and \cite{Forogione20}, two different specifications are applied to the MPC controller and both behaviors are presented to the user. The user chooses the preferred one. To repeat this, the optimal specification for the user is obtained.
In addition, \cite{Nii2023} proposes a method to optimize the specification of the optimal control system. The specification is updated by the rating provided by the user.
Moreover, \cite{Zhou21} proposes an algorithm to generate a path that both satisfies the constraints and maximizes the user's satisfaction.

Since \cite{Zhu2021}, \cite{Maccarini22}, and \cite{Forogione20} require repeatedly collecting user feedback to optimize the specification, it imposes a burden on the user since the user has to answer the survey one by one. 
Another drawback can be seen in \cite{Nii2023}: it expects a questionnaire format as the survey from the user. The survey answers are collected every few days, so the low update frequency might be the problem.

In this paper, we attempt to overcome the drawbacks existing in the previous studies on personalization. To reduce the burden on the user and collect data in high frequency, we use the chat in natural languages to communicate with the system. We introduce the novel framework ``ChatMPC'', for personalizing MPC by collecting chat in natural language from the user.

\begin{figure}[b]
    \centering
    \includegraphics[width=1\linewidth]{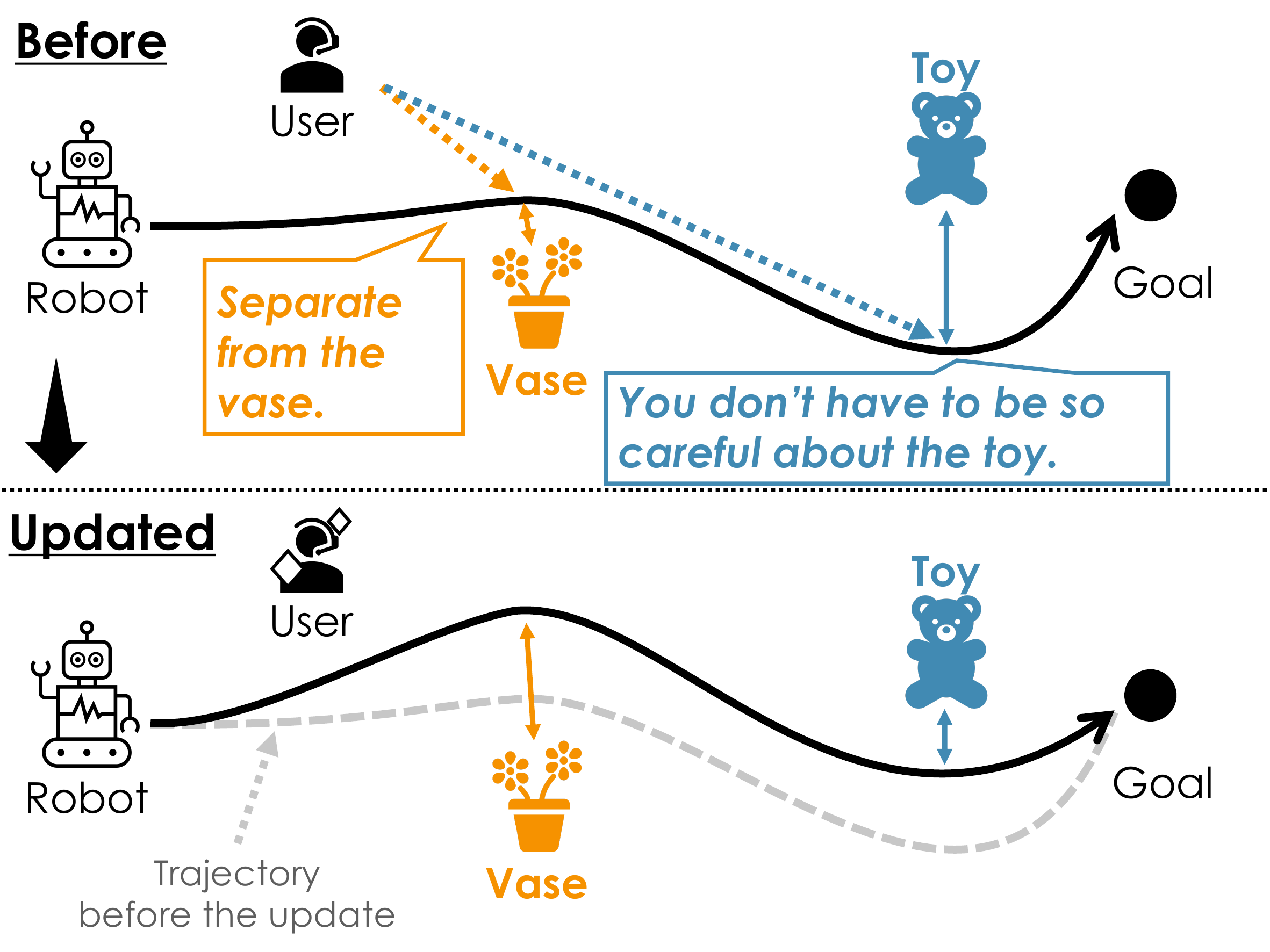}
    \caption{One example of the use case of ChatMPC}
    \label{F.Introduction.Concept}
\end{figure}

\figref{F.Introduction.Concept} illustrates a typical use case of the ChatMPC. In this figure, we consider an autonomous traveling robot, running toward the goal point while avoiding obstacles, \vase and \toy. 
This robot adjusts its behavior according to instructions in natural language by the user. 
For instance, if the user instructs the robot to keep the vase at enough distance, the robot runs on a path away from the vase. Next, if the user instructs that the robot does not have to keep away from the toy, the robot generates a path not far enough away from the toy and prioritizes arrival at the goal point.
In this way, the control system is updated by user instructions, and the behavior of the robot adapts to the user's preferences.

There have been several studies on cooperation between natural languages and robots. For example, \cite{Gubbi21}, \cite{Tellex11}, and \cite{Branavan09} propose a method to generate the robot's preferred actions based on sentences in natural language. A language model that translates sentences in natural language into a Python code is used in \cite{Gubbi21}, and reinforcement learning is used in \cite{Tellex11} and \cite{Branavan09}.
Other works \cite{Howard2014} and \cite{Park19} propose a method to translate sentences in natural language into constraints enabling a control system to achieve the purpose of controlling. 
Additionally, \cite{Yu23} and \cite{Tang23} utilize a large language model (LLM) to manipulate a robot. In \cite{Yu23}, LLM is used to translate sentences into reward functions enabling a robot to achieve the purpose of controlling, and in \cite{Tang23}, LLM is used to generate a quadrupedal locomotion's desired foot contact patterns.
One can see that these methods are to generate a control objective or a trajectory according to natural language instructions.

ChatMPC introduced in this paper is intended to determine the specification of the controller. While control objectives and trajectories are completed once a single control trial is finished, the specifications of the controller function as persistently effective characteristics even after a single control trial is finished. By personalizing the controller's specifications, its effectiveness is consistently demonstrated regardless of the operating environment.

\section{ChatMPC}

\subsection{Overview}

\begin{figure}[t]
    \centering
    \includegraphics[width=1\linewidth]{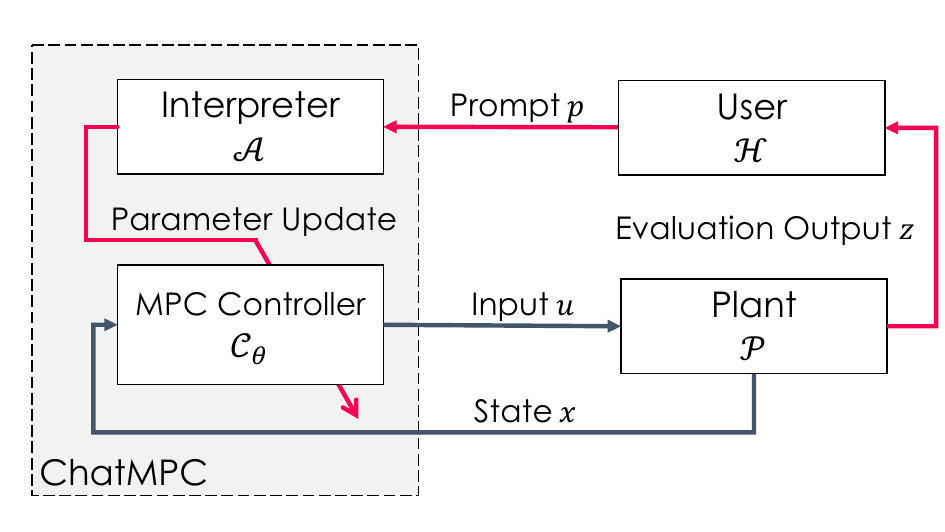}
    \caption{Overall structure of ChatMPC}
    \label{F.ChatMPC.Overview}
\end{figure}
\figref{F.ChatMPC.Overview} shows the overall structure of ChatMPC. In this figure, $\Plant$ is a plant, $\Controller$ is an MPC controller, $\User$ is a user, and $\Interpreter$ is called \textit{interpreter}.
Symbol $\theta$ represents a parameter associated with the MPC controller and is externally adjustable. In \figref{F.ChatMPC.Overview}, the gray region enclosed by the black line represents the ChatMPC.

The black line in \figref{F.ChatMPC.Overview} represents a typical loop of feedback control, which is called the \textit{control loop}. In the control loop, the MPC controller determines the control input $u$ based on the state of the plant $x$. 
The red line is called the \textit{personalization loop}, which is an original in this paper. In the personalization loop, the user $\User$ receives the evaluation output $z$ from the plant $\Plant$, and the user provides natural language strings to the interpreter $\Interpreter$ as opinions and advice regarding the evaluation output $z$. The natural language string is called the prompt $p$ in this paper. Then, the interpreter $\Interpreter$ updates the parameter $\theta$ of the MPC controller $\Controller$. Through the personalization loop, the specification of the MPC controller changes, enabling the behavior of the MPC controller to match the user's preferences. 

Note that the time scale is different between the two loops; the control loop (black line) and the personalization loop (red line). For instance, assuming the autonomous electric wheelchair, the control loop runs once a few milliseconds, while the personalization loop runs once a few seconds or minutes. 
In this paper, the time step of the control loop and the personalization loop are denoted by $k$ and $\tau$ respectively.

In the following subsections, we provide detailed explanations of each component of ChatMPC. We describe the MPC controller in \ref{S.ChatMPC.MPCController}, we present an example of the MPC controller and the interpreter in \ref{S.ChatMPC.SafetyConstraintInMPC} and \ref{S.ChatMPC.InternalStructureOfInterpreter}, respectively.

\subsection{MPC Controller}
\label{S.ChatMPC.MPCController}

First, we assume a control system that consists of the plant $\Plant$ and the MPC controller $\Controller$. The plant is described by the discrete-time state-space representation:
\begin{equation}
    \Plant :
    \begin{cases}
        x(k+1) = f(x(k),u(k)),\\
        z(k) = g(x(k),u(k)),
    \end{cases}
    \label{E.ChatMPC.Plant}
\end{equation} 
where $x\in\xField$ and $u\in\uField$ are the plant state and the plant input, respectively, and $z\in\zField$ is the evaluation output. Symbol $f:\xField\times\uField\to\xField$ and $g:\xField\times\uField\to\zField$ represent the plant state and the evaluation output mapping, respectively. 

Suppose the MPC controller with a time tick of $\Delta t$, a prediction horizon of $N_p$, and a control horizon of $1$. At each time step $k$, the controller calculates the optimal input $u(k|k)$, where $u(k+i|k)$ represents the control input of the future time step $k+i$ calculated in the time step $k$.
For simplicity, we introduce the state sequence $X(k)\InRealField{\XDim}$ and the input sequence $U(k)\InRealField{\UDim}$ are defined as follows:
\begin{align*}
    & U(k) = [ u(k|k)^\top~\cdots~u(k+N_p-1|k)^\top ]^\top \\
    & X(k) = [ x(k+1|k)^\top~\cdots~x(k+N_p|k)^\top ]^\top
    .
\end{align*}
In this case, the optimal input $u(k|k)$ is obtained by solving the following optimization problem:
\begin{align}
    \Controller:\begin{cases}\begin{split}
        \min_{U(k)}~
            & J_\theta(X(k),U(k))\\
        \text{s.t.}~
            & x(k|k)=x(k), \\
            & x(k+i+1|k)=f(x(k+i|k),u(k+i|k)), \\
                & \qquad \forall i \InRange{0}{N_p-1} \\
            & X(k) \in \XSet, \\
            & U(k) \in \USet
    ,
    \end{split}\end{cases}
    \label{E.ChatMPC.MPCOptimizer}
\end{align} 
where $\theta\in\thetaField$ is the adjustable parameter, $J_\theta(X,U):\XField\times\UField\to\RealField{}$ is the cost function, $\XSet\subseteq\XField$, $\USet\subseteq\UField$ are sets of the allowable state sequence and the input sequence, respectively.

It should be emphasized here that symbols $J_\theta$, $\XSet$, and $\USet$ are parameterized by $\theta$, and $\theta$ is updated through the personalization loop.

\subsection{Safety Constraint in MPC}
\label{S.ChatMPC.SafetyConstraintInMPC}

In the previous subsection, we described the basic structure of an MPC controller. In this subsection, we present a specialization of the MPC controller, named ``MPC-CBF''. MPC-CBF is proposed in \cite{Zeng21} which incorporates control barrier functions (CBF) into the optimization problem of the MPC controller \eqref{E.ChatMPC.MPCOptimizer}.

CBF is a function $h$ that satisfies the following inequality \cite{Ames19}:
\begin{equation} 
    \exists u ~
    \text{s.t.} ~
    \dot h(x) = \frac{\partial h(x)}{\partial x}f(x,u)
    \ge -\gamma(h(x))
    ,
    \label{E.ChatMPC.CBFFullVersion}
\end{equation} 
where $f$ is a function that characterizes the plant dynamics, i.e., $f$ satisfies $\dot x=f(x,u)$, and $\gamma(h)$ is a continuously differentiable strictly increasing function with $\gamma(0)=0$ and $\gamma(h)\to\infty,h\to\infty$.

Next, we convert \eqref{E.ChatMPC.CBFFullVersion} to a discrete-time, simplified form that aligns with the notation of MPC. For simplicity, we introduce $h_i=h(x(k+i|k))$ and $\Delta h_i=h_{i+1}-h_i$. Then, we have \eqref{E.ChatMPC.CBFFullVersion} as:
\begin{equation} 
    \Delta h_i(x) \ge -\gamma h_i(x)
    ,
    \label{E.ChatMPC.CBFEasyVersion}
\end{equation}
where $\gamma$ is the positive constant.
Now, the CBF and the state $x$ hold the following property:
If the initial state at time $0$ satisfies $h(x(0))\ge0$, and for all times $k=0,1,\ldots$ the inequality \eqref{E.ChatMPC.CBFEasyVersion} holds, $x$ continues to satisfy $h(x(t))\ge0$ for all times $k\in\{0,1,\ldots\}$. 

MPC-CBF utilizes the property and incorporates the CBF inequality \eqref{E.ChatMPC.CBFEasyVersion} as a constraint on the state sequence $X(k)$. Specifically, we describe the set of allowable state sequence $\XSet$ as follows:
\begin{equation}
    \XSet = \left\{
        X(k) \midmid
        \Delta h_i \ge -\gamma h_i
        ,~ \forall i \InRange{1}{N_p}
    \right\}
    .
    \label{E.ChatMPC.XConstraintByCBF}
\end{equation} 

\subsection{Internal Structure of Interpreter}
\label{S.ChatMPC.InternalStructureOfInterpreter}

The interpreter $\Interpreter$ is an important component of the personalization loop in ChatMPC. The interpreter has the role of adjusting the parameter $\theta$ according to the content of the provided prompt $p$. 

\begin{figure}[t]
    \centering
    \includegraphics[width=1\linewidth]{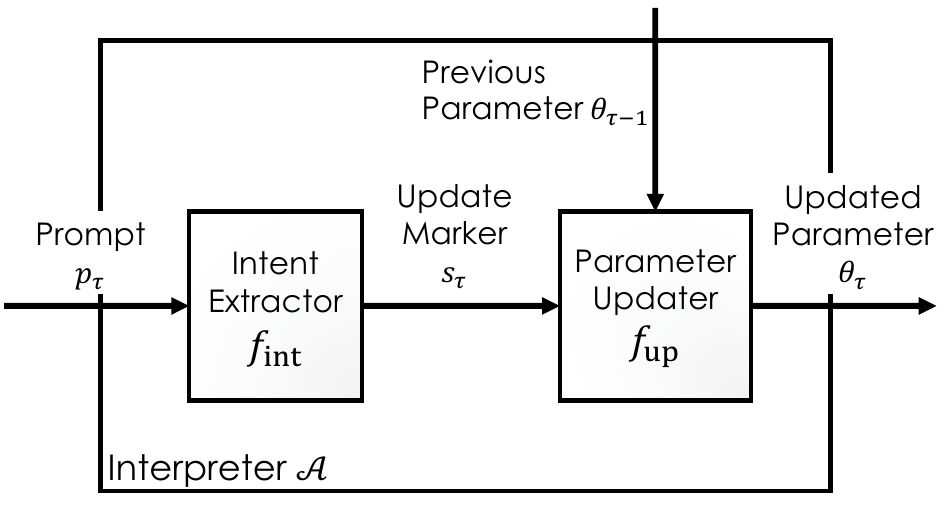}
    \caption{Structure of interpreter}
    \label{F.ChatMPC.Interpreter}
\end{figure}

\figref{F.ChatMPC.Interpreter} shows the structure of the interpreter.
Denoting the iteration count in the personalization loop as $\tau\in\{1,2,\ldots\}$, we assume that the interpreter $\Interpreter$ receives the $\tau$-th prompt $p_\tau$ from the user $\User$.
The interpreter $\Interpreter$ consists of an intent extractor $\fint$ and a parameter updater $\fup$. 

First, the intent extractor $\fint$ analyzes the content of the prompt $p_\tau$ and outputs $s_\tau\InRealField{\thetaDim}$ based on the intent of the prompt. Symbol $s_\tau$ is called the update marker, which provides the information about which element of parameter $\theta_{\tau-1}$ should be updated. The update marker can be expressed as follows:
\begin{equation}
    s_\tau = \fint(p_\tau) \in\{-1,0,+1\}^n \subset\RealField{\thetaDim}
    \label{E.ChatMPC.ParameterUpdateMarker}
\end{equation} 
and a non-zero element indicates that the corresponding element of the parameter should be updated.

Next, the parameter updater $\fup$ calculates the updated parameter $\theta_\tau$ based on the update marker $s_\tau$ and the previous parameter $\theta_{\tau-1}$. It can be expressed as follows:
\begin{equation}
    \theta_\tau = \fup(s_\tau,\theta_{\tau-1}) 
    = \mathrm{pow}(d,s_\tau) \odot \theta_{\tau-1}
    ,
    \label{E.ChatMPC.ParameterUpdater}
\end{equation} 
where $\mathrm{pow}(a,b)$ represents the element-wise power, i.e., $\mathrm{pow}(a,b)=[a_1^{b_1}~\cdots~a_n^{b_n}]^\top$,
and $d\InRealField{\thetaDim}$ is a constant vector that specifies the increase/decrease of the parameter, which is called the update constant.
In this way, the parameter is updated according to the intent of the prompt.

For the implementation of $\fint$, it can be effective to incorporate a natural language model such as Sentence BERT model\cite{Nils19}. Details of the implementation are given in Subsection \ref{S.NE.Interpreter}.
In addition, the formulations of $\fup$ can be different from \eqref{E.ChatMPC.ParameterUpdater} like $\fup(x_\tau,\theta_{\tau-1})=\theta_{\tau-1}+d\odot s_\tau$.

\section{Numerical Experiment}

In this numerical experiment, we build a simulation of an autonomous cleaning robot equipped with ChatMPC. In the simulation environment, there are multiple obstacles, and the robot is required to navigate to the goal point while avoiding these obstacles. There are two types of obstacles; $\vase$ and $\toy$. In this simulation, we are able to personalize the avoidance behavior for each type of obstacle.
\footnote{The code for the numerical experiment is available on \texttt{https://github.com/Mya-Mya/ChatMPC}.}

\subsection{Obstacle}

We assume that we have $N$ obstacles in a simulation environment and let $j\InRange{1}{N}$ denote the index of obstacle $j$.
We use $m(j)\in\{\vase,\toy\}$, $(x_1\IdxJ,x_2\IdxJ)$, and $R\IdxJ$ to represent the type, position, and safety margin of obstacle $j$, respectively.

\subsection{Plant Model}

Let $(x_1,x_2)$ denote the position of the robot, and let $v_1,v_2$ denote the velocities in the $x_1$ and $x_2$ directions, respectively. We define the state of the robot as $x=[x_1~x_2~v_1~v_2]^\top$. Then, the state equation of the robot is given as follows:
\begin{equation}
    x(k+1)=Ax(k)+Bu(k)
    ,
    \label{E.NE.Plant}
\end{equation} 
where $u$ represents the input applied to the robot and the robot operates with a time interval $\Delta t=0.2\Unit{s}$. The matrices $A$ and $B$ are defined as: 
\begin{equation}
A=
\begin{bmatrix}
    1&0&\Delta t&0\\
    0&1&0&\Delta t\\
    0&0&1&0\\
    0&0&0&1
\end{bmatrix},\\
B=
\begin{bmatrix}
    0&0 \\
    0&0 \\
    \Delta t&0\\
    0&\Delta t
\end{bmatrix}
.
\end{equation} 

\subsection{Control Problem and Parameter}

The control objective in this numerical experiment is to navigate the robot, starting from the initial state $x_0$, to the goal position $(0,0)$ while avoiding obstacles.
In order to personalize the avoidance behavior for each type of obstacle, we select the constants used within CBF as the adjustable parameter $\theta$.

We use MPC-CBF presented in Subsection \ref{S.ChatMPC.SafetyConstraintInMPC} as the MPC controller $\Controller$, with the prediction horizon of $N_p=8$, and the time tick of $\Delta t=0.2\Unit{s}$.

The cost function $J_\theta(X(k),U(k))$ is defined as follows:
\begin{align}\begin{split}
    &J(X(k),U(k)) \\
    & = \sum_{i=1}^{N_p} l(x(k+i|k),u(k+i|k))
    ~ +\phi(x(k+N_p|k)
    ,
    \label{E.NE.Cost}
\end{split}\end{align} 
where
\begin{align}
    & l(x,u) = x^\top Q x + u^\top R u, \\
    & \phi(x) = x^\top P x 
\end{align} 
with $Q=\Diag{1,1,1,1}$,$R=\Diag{1,1}$,$P=\Diag{100,100,100,100}$.
The set of input sequence $U(k)$ is defined as follows:
\begin{align}\begin{split}
    \mathcal{U} = \{ U(k) \mid~
        & u(k+i|k)\in[-1,1]^2, \\
        & \forall i \InRange{0}{N_p-1} 
    \}
    ,
\end{split}\end{align} 
and the constraint on the input sequence can be expressed as $U(k)\in\mathcal{U}$. 
Note that the cost function $J$ and input constraint $\mathcal{U}$ are independent of the adjustable parameter $\theta$.

Next, we use CBF to construct the constraint on the state sequence.
First, we define a function $h\IdxJ(x)$ for each obstacle $j\InRange{1}{N}$ on the simulation environment as follows:
\begin{equation}
    h\IdxJ(x)
    = (x_1-x_1\IdxJ)^2 + (x_2-x_2\IdxJ)^2
    - (R\IdxJ)^2
    .
    \label{E.NE.CBFForEachObstacle}
\end{equation} 
This equation only allows the robot to navigate outside the safety margin $R\IdxJ$ of obstacle $j$.
Next, using this function $h\IdxJ(x)$, we define a barrier inequality for the states $x(k+i|k)$ and $x(k+i+1|k),~i\InRange{0}{N_p-1}$.
\begin{equation}
     \Delta h_i\IdxJ \ge -\gamma_{m(j)} h_i\IdxJ
     ,
    \label{E.NE.xConstraintForEachObstacle}
\end{equation} 
where $h\IdxJ(x(k+i|k))=h_i\IdxJ$, $\Delta h_i\IdxJ=h_{i+1}\IdxJ-h_i\IdxJ$.
Note that the inequality \eqref{E.NE.xConstraintForEachObstacle} depends on the adjustable parameter $\gamma_{m(j)}$. 
$\gamma_{m(j)}$ represents a value of the parameter depending on the type of the obstacle $j$. If the type of the obstacle is $\vase$, $\gamma_{m(j)}=\gamma_\vase$ and if the type of the obstacle is $\toy$, $\gamma_{m(j)}=\gamma_\toy$. Both $\gamma_\vase$ and $\gamma_\toy$ are the values of the parameter and can be updated in the personalization loop.
Finally, by collecting \eqref{E.NE.xConstraintForEachObstacle} for all obstacles $j,~j\InRange{1}{N}$ and all time steps $i,~i\InRange{0}{N_p-1}$, we obtain the set of the state sequence $\XSet$ as follows:
\begin{align}\begin{split}
    \XSet = \{
        X(k) \mid~
            & \Delta h_i\IdxJ \ge -\gamma_{m(j)} h_i\IdxJ
         \\
            & \forall j\InRange{1}{N} 
            , \forall i\InRange{0}{N_p-1} 
    \}
    .
\end{split}\end{align} 
Therefore, the constraint on $X(k)$ is expressed as $X(k)\in\XSet$.

In this numerical experiment, there are two values as the adjustable parameter $\theta$ as follows:
\begin{equation}
    \theta = [\gamma_\vase~\gamma_\toy]^\top
    .
    \label{E.NE.Parameters}
\end{equation} 
These values parameterize the constraint on the state sequence $\XSet$.

\subsection{Interpreter}
\label{S.NE.Interpreter}

\begin{figure*}
    \centering
    \includegraphics[width=1\linewidth]{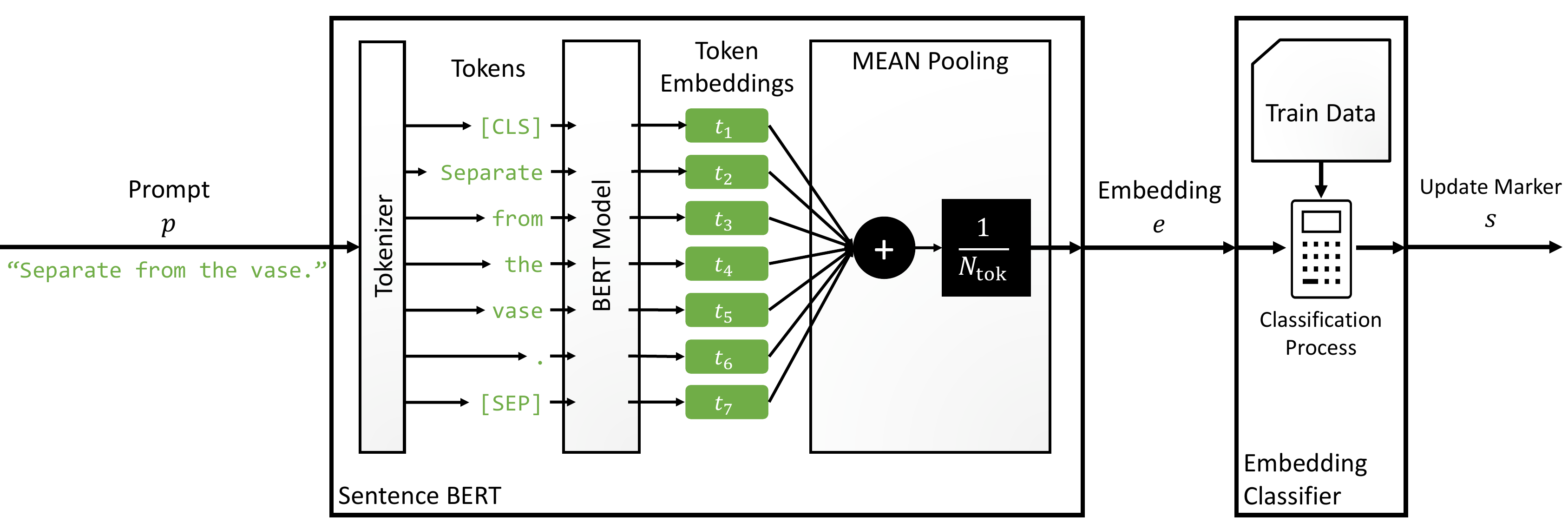}
    \caption{%
    Structure of the intent extractor:
    Sentence BERT consists of Tokenizer, BERT model, and MEAN Pooling. First, the Tokenizer tokenizes the provided prompt $p$ into $N_\mathrm{tok}$ tokens. Next, the BERT model processes the tokens and outputs the token embeddings $t_j\InRealField{\eDim},~j\InRange{1}{N_\mathrm{tok}}$. Finally, the Mean Pooling calculates the average of the token embedding and outputs the embedding $e\InRealField{\eDim}$.
    The embedding classifier classifies the provided embedding $e$ into the update marker $s$ using the train data.}
    \label{F.NE.IntentExtractorStructure}
\end{figure*}

The internal structure of the intent extractor $\fint$ is shown in \figref{F.NE.IntentExtractorStructure}. The intent extractor consists of two parts; Sentence BERT and an embedding classifier.
Sentence BERT is a language model that takes a sentence and outputs an embedding $e\InRealField{\eDim}$ that represents the contextual information of the sentence \cite{Nils19}. In this numerical experiment, we used \texttt{deepset/sentence\_bert} by deepset as the pre-trained weight of the Sentence BERT model, available at \cite{Deepset21}.
Then, the embedding $e$ is classified into multiple classes by the embedding classifier. An update marker $s$ is defined for each class and serves as the final output of $\fint$. 

In this numerical experiment, the intent extractor $\fint$ classifies the prompt $p$ into four classes. We prepare some example prompts and update markers for the train data of the embedding classifier. The example prompts are shown in \tableref{T.NE.TrainData}.
\begin{table*}[t]
    \centering
    \caption{Prompts and update markers for the train data}
    \begin{tabular}{l|c}
        \hline
        Example prompts & Update marker $s$  \\
         \hline

         \begin{tabular}{l}
            \PromptContent{Can you separate from the vase?}\\
            \PromptContent{Please separate from the vase.}\\
            \PromptContent{It is too close to the vase.}\\
            \PromptContent{Too close to the vase.}\\
            \PromptContent{You are too closing to the vase}\\
        \end{tabular}
        & $[-1~0]^\top$ \\
        \hline

        \begin{tabular}{l}
            \PromptContent{Can you approach to the vase?}\\
            \PromptContent{Please approach to the vase.}\\
            \PromptContent{You do not need to care about the vase.}\\
            \PromptContent{You do not need to be careful about the vase.}\\
            \PromptContent{You do not have to care about the vase so much.}\\
        \end{tabular}
        & $[+1~0]^\top$ \\
        \hline

        \begin{tabular}{l}
            \PromptContent{Can you separate from the toy?}\\
            \PromptContent{Please separate from the toy.}\\
            \PromptContent{It is too close to the toy.}\\
            \PromptContent{Too close to the toy.}\\
            \PromptContent{You are too closing to the toy}\\
        \end{tabular}
        & $[0~-1]^\top$ \\
        \hline

        \begin{tabular}{l}
            \PromptContent{Can you approach to the toy?}\\
            \PromptContent{Please approach to the toy.}\\
            \PromptContent{You do not need to care about the toy.}\\
            \PromptContent{You do not need to be careful about the toy.}\\
            \PromptContent{You do not have to care about the toy so much.}\\
        \end{tabular}
        & $[0~+1]^\top$ \\
         \hline
    \end{tabular}
    \label{T.NE.TrainData}
\end{table*}

The constant $d$ used in the parameter updater $\fup$ is set to $d=[2~2]^\top$.

\subsection{Procedure}

The initial values of the parameter are set to $\theta_0=[0.4~0.4]^\top$ and two environments, called A and B, are prepared. 
In Environment A, the $\vase$ is placed at $(-1,-3)$, the $\toy$ is placed at $(-3,-1)$, and the initial state of the robot is $x_0=[-5~-5~0~0]^\top$. 
In Environment B, the $\vase$ is placed at $(-1,-4)$ and $(-1,-2)$, the $\toy$ is placed at $(1.5,-3)$, and the initial state of the robot is $x_0=[0~-10~0~0]^\top$.
The safety margin for all obstacles is set to $R\IdxJ=R=0.5$.

In each environment, the robot runs from the start point to the goal point three times, Trial 1, Trial 2, and Trial 3 respectively. 
Trial 1 is performed without providing any prompt. In Trial 2 and Trial 3, we provide the prompts as shown in \tableref{T.NE.ProvidedPrompts}.
\begin{table*}[t]
    \centering
    \caption{Prompt provided in each trial}
    \begin{tabular}{c|l}
        \hline
         Trial number & Provided prompt \\
         \hline
         Trial 1 & \textit{no prompt}  \\
         Trial 2 & $p_1=$\PromptContent{Separate from the vase.}\\
         Trial 3 & $p_2=$\PromptContent{You don't have to be so careful about the toy.}\\
         \hline
    \end{tabular}
    \label{T.NE.ProvidedPrompts}
\end{table*}

\subsection{Result}

\begin{figure}[t]
    \centering
    \includegraphics[width=1\linewidth]{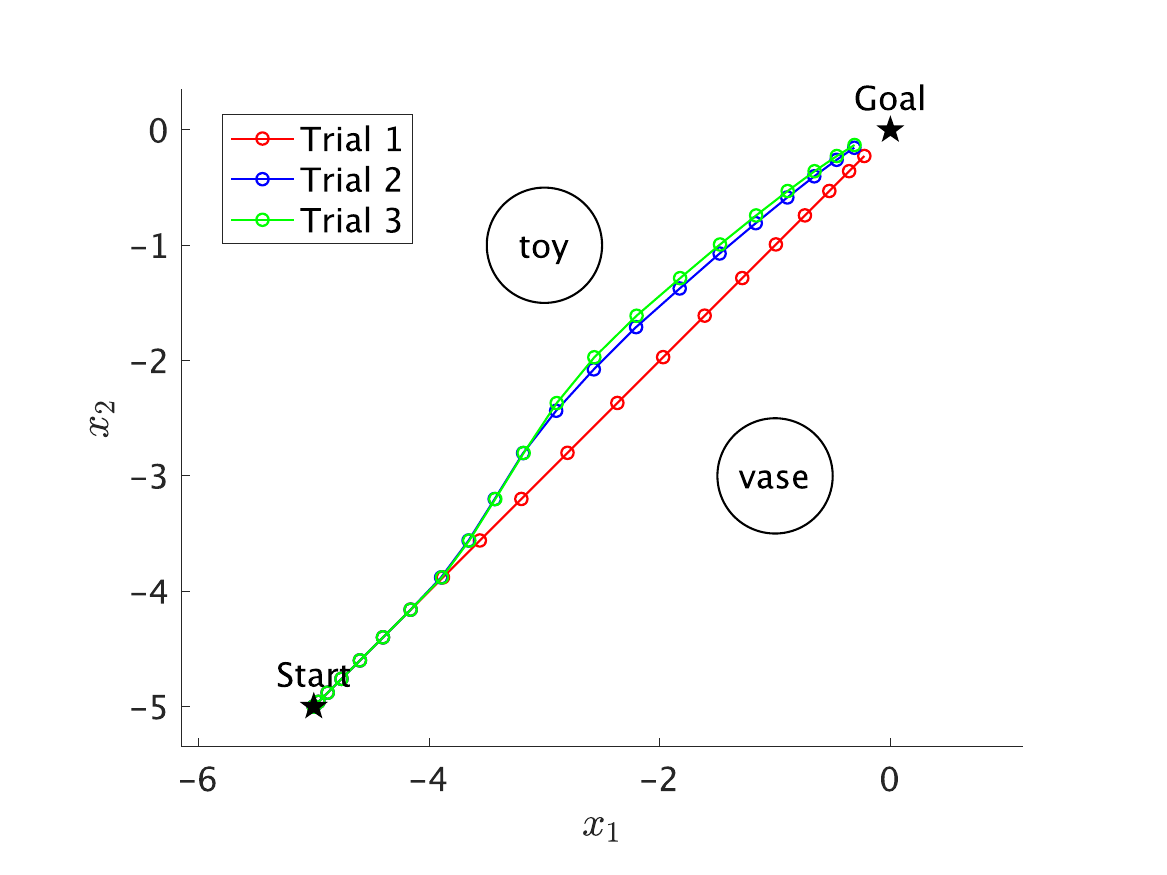}
    \caption{Robot trajectory in each trial in environment A}
    \label{F.NE.EnvA}
\end{figure}

\begin{figure}[t]
    \centering
    \includegraphics[width=1\linewidth]{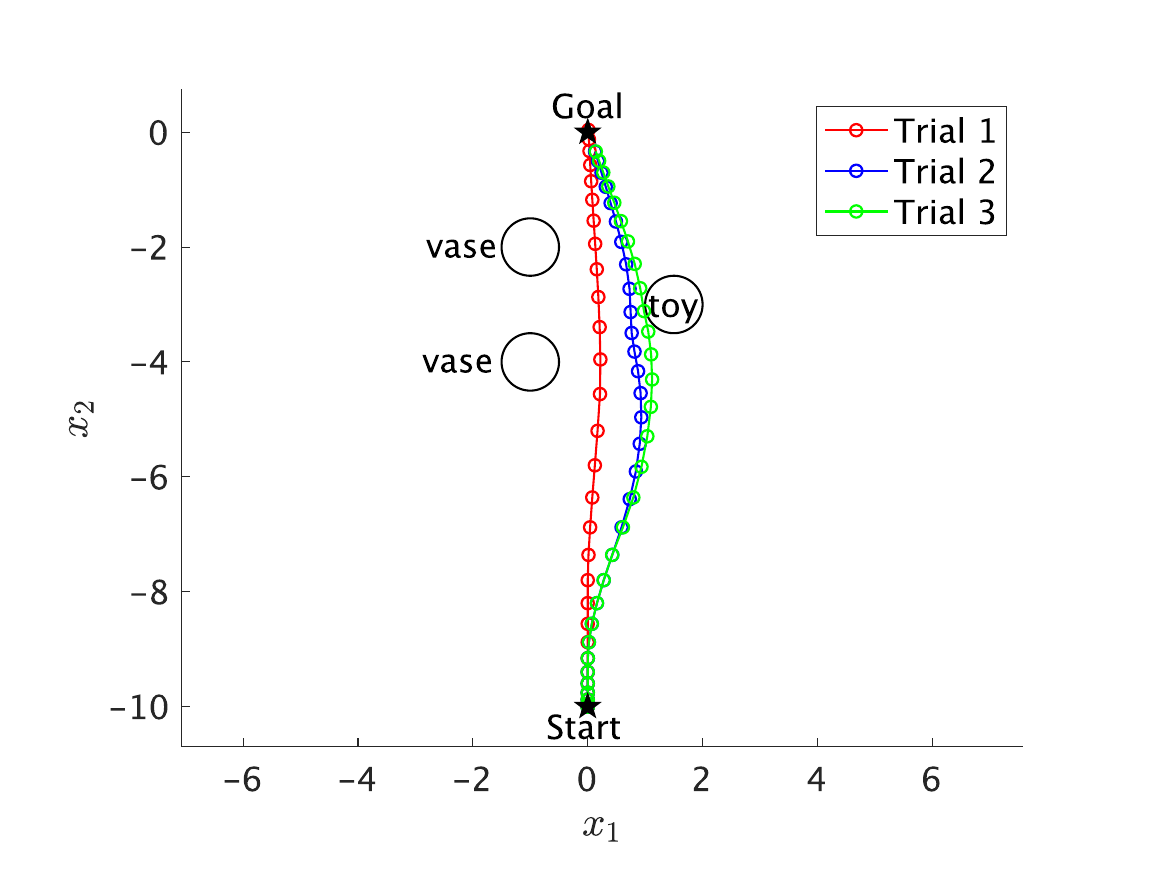}
    \caption{Robot trajectory in each trial in environment B}
    \label{F.NE.EnvB}
\end{figure}

The robot's trajectories in Environment A are shown in \figref{F.NE.EnvA}, and the robot's trajectories in Environment B are shown in \figref{F.NE.EnvB}.
With each trial, the robot's trajectory gradually moves away from the $\vase$ and towards the $\toy$. The specification of the MPC controller is updated based on the prompts provided, resulting in more preferable behavior. 
Additionally, the behavior changes that are consistent across all environments imply that the personalization of the control system is applied consistently regardless of the operating environment.

\section{Conclusion}

We proposed ChatMPC, a personalization framework.
ChatMPC updates the specification of the MPC controller based on natural language prompts provided and aims to provide behavior that is preferable for the user.
By personalizing the specifications of the controller, its efficacy is consistently demonstrated regardless of the surrounding environment.

In the numerical experiment, we equipped a robot with ChatMPC and simulated it in multiple environments while providing prompts. The result shows that the robot effectively adapted its behavior according to the intentions of the prompts.

We emphasize that the interpreter, which is the core component in ChatMPC and given in Subsection \ref{S.ChatMPC.InternalStructureOfInterpreter}, is applicable to any control systems that are based on optimization problems other than MPC. 
In addition, we will address the analysis of the personalization loop in ChatMPC, including its convergence analysis under some assumptions on user models.


\section*{Acknowledgement}
The authors would like to thank for Prof.~J.~M.~Maestre for his valuable comments on this work.

This work was supported by Grant-in-Aid for Scientific Research (B), No.~20H02173 from JSPS.

\bibliographystyle{IEEEtran}
\bibliography{references}

\begin{thebibliography}{10}
\providecommand{\url}[1]{#1}
\csname url@samestyle\endcsname
\providecommand{\newblock}{\relax}
\providecommand{\bibinfo}[2]{#2}
\providecommand{\BIBentrySTDinterwordspacing}{\spaceskip=0pt\relax}
\providecommand{\BIBentryALTinterwordstretchfactor}{4}
\providecommand{\BIBentryALTinterwordspacing}{\spaceskip=\fontdimen2\font plus
\BIBentryALTinterwordstretchfactor\fontdimen3\font minus
  \fontdimen4\font\relax}
\providecommand{\BIBforeignlanguage}[2]{{%
\expandafter\ifx\csname l@#1\endcsname\relax
\typeout{** WARNING: IEEEtran.bst: No hyphenation pattern has been}%
\typeout{** loaded for the language `#1'. Using the pattern for}%
\typeout{** the default language instead.}%
\else
\language=\csname l@#1\endcsname
\fi
#2}}
\providecommand{\BIBdecl}{\relax}
\BIBdecl

\bibitem{Bujarbaruah18}
M.~Bujarbaruah, X.~Zhang, and F.~Borrelli, ``{Adaptive MPC with Chance
  Constraints for FIR Systems},'' in \emph{2018 Annual American Control
  Conference (ACC)}, 2018, pp. 2312--2317.

\bibitem{Lorenzen19}
M.~Lorenzen, M.~Cannon, and F.~Allg{\"o}wer, ``{Robust MPC with recursive model
  update},'' \emph{Automatica}, vol. 103, pp. 461--471, 2019.

\bibitem{Tanaskovic13}
M.~Tanaskovic, L.~Fagiano, R.~Smith, P.~Goulart, and M.~Morari, ``{Adaptive
  model predictive control for constrained linear systems},'' in \emph{2013
  European Control Conference (ECC)}, 2013, pp. 382--387.

\bibitem{Liang22}
Z.~Liang and J.~K.~C. Lo, ``{An Iterative Method to Learn a Linear Control
  Barrier Function},'' \emph{ArXiv}, vol. abs/2211.09854, 2022.

\bibitem{Parwana2022}
H.~Parwana, A.~Mustafa, and D.~Panagou, ``{Trust-based Rate-Tunable Control
  Barrier Functions for Non-Cooperative Multi-Agent Systems},'' in \emph{2022
  IEEE 61st Conference on Decision and Control (CDC)}, 2022, pp. 2222--2229.

\bibitem{Zhu2021}
M.~Zhu, A.~Bemporad, and D.~Piga, ``{Preference-based MPC calibration},'' in
  \emph{2021 European Control Conference (ECC)}, 2021, pp. 638--645.

\bibitem{Maccarini22}
M.~Maccarini, F.~Pura, D.~Piga, L.~Roveda, L.~Mantovani, and F.~Braghin,
  ``{Preference-Based Optimization of a Human-Robot Collaborative
  Controller},'' \emph{IFAC-PapersOnLine}, vol.~55, no.~38, pp. 7--12, 2022,
  13th IFAC Symposium on Robot Control SYROCO 2022.

\bibitem{Forogione20}
M.~Forgione, D.~Piga, and A.~Bemporad, ``{Efficient Calibration of Embedded
  MPC},'' \emph{IFAC-PapersOnLine}, vol.~53, no.~2, pp. 5189--5194, 2020, 21st
  IFAC World Congress.

\bibitem{Nii2023}
T.~Nii and M.~Inoue, ``{Personalization of Control Systems by Policy Update
  With Improving User Trust},'' \emph{IEEE Control Systems Letters}, vol.~7,
  pp. 889--894, 2023.

\bibitem{Zhou21}
Y.~Zhou, Y.~Zhang, X.~Luo, and M.~M. Zavlanos, ``{Human-in-the-Loop Robot
  Planning with Non-Contextual Bandit Feedback},'' in \emph{2021 60th IEEE
  Conference on Decision and Control (CDC)}, 2021, pp. 2848--2853.

\bibitem{Gubbi21}
S.~G. Venkatesh, R.~Upadrashta, and B.~Amrutur, ``{Translating Natural Language
  Instructions to Computer Programs for Robot Manipulation},'' in \emph{2021
  IEEE/RSJ International Conference on Intelligent Robots and Systems
  (IROS)}.\hskip 1em plus 0.5em minus 0.4em\relax IEEE Press, 2021, pp.
  1919--1926.

\bibitem{Tellex11}
S.~Tellex, T.~Kollar, S.~Dickerson, M.~R. Walter, A.~G. Banerjee, S.~Teller,
  and N.~Roy, ``{Understanding Natural Language Commands for Robotic Navigation
  and Mobile Manipulation},'' in \emph{Proceedings of the Twenty-Fifth AAAI
  Conference on Artificial Intelligence}, ser. AAAI'11.\hskip 1em plus 0.5em
  minus 0.4em\relax AAAI Press, 2011, pp. 1507--1514.

\bibitem{Branavan09}
S.~R.~K. Branavan, H.~Chen, L.~S. Zettlemoyer, and R.~Barzilay,
  ``{Reinforcement Learning for Mapping Instructions to Actions},'' in
  \emph{Proceedings of the Joint Conference of the 47th Annual Meeting of the
  ACL and the 4th International Joint Conference on Natural Language Processing
  of the AFNLP: Volume 1 - Volume 1}, ser. ACL '09.\hskip 1em plus 0.5em minus
  0.4em\relax USA: Association for Computational Linguistics, 2009, pp. 82--90.

\bibitem{Howard2014}
T.~M. Howard, S.~Tellex, and N.~Roy, ``{A natural language planner interface
  for mobile manipulators},'' in \emph{2014 IEEE International Conference on
  Robotics and Automation (ICRA)}, 2014, pp. 6652--6659.

\bibitem{Park19}
J.~S. Park, B.~Jia, M.~Bansal, and D.~Manocha, ``{Efficient Generation of
  Motion Plans from Attribute-Based Natural Language Instructions Using Dynamic
  Constraint Mapping},'' in \emph{2019 International Conference on Robotics and
  Automation (ICRA)}.\hskip 1em plus 0.5em minus 0.4em\relax IEEE Press, 2019,
  pp. 6964--6971.

\bibitem{Yu23}
W.~Yu, N.~Gileadi, C.~Fu, S.~Kirmani, K.-H. Lee, M.~G. Arenas, H.-T.~L. Chiang,
  T.~Erez, L.~Hasenclever, J.~Humplik, B.~Ichter, T.~Xiao, P.~Xu, A.~Zeng,
  T.~Zhang, N.~Heess, D.~Sadigh, J.~Tan, Y.~Tassa, and F.~Xia, ``{Language to
  Rewards for Robotic Skill Synthesis},'' 2023.

\bibitem{Tang23}
Y.~Tang, W.~Yu, J.~Tan, H.~Zen, A.~Faust, and T.~Harada, ``{SayTap: Language to
  Quadrupedal Locomotion},'' 2023.

\bibitem{Zeng21}
J.~Zeng, B.~Zhang, and K.~Sreenath, ``{Safety-Critical Model Predictive Control
  with Discrete-Time Control Barrier Function},'' in \emph{2021 American
  Control Conference (ACC)}, 2021, pp. 3882--3889.

\bibitem{Ames19}
A.~D. Ames, S.~Coogan, M.~Egerstedt, G.~Notomista, K.~Sreenath, and P.~Tabuada,
  ``{Control Barrier Functions: Theory and Applications},'' in \emph{2019 18th
  European Control Conference (ECC)}, 2019, pp. 3420--3431.

\bibitem{Nils19}
N.~Reimers and I.~Gurevych, ``{Sentence-BERT: Sentence Embeddings using Siamese
  BERT-Networks},'' \emph{CoRR}, vol. abs/1908.10084, 2019.

\bibitem{Deepset21}
\BIBentryALTinterwordspacing
deepset. (2021, 5) {Hugging Face: deepset/sentence\_bert}. [Online]. Available:
  \url{https://huggingface.co/deepset/sentence_bert}
\BIBentrySTDinterwordspacing

\end{thebibliography}

\end{document}